\documentclass[aps,amssymb,prl,twocolumn]{revtex4}
\usepackage{graphicx}

\def\d {{\rm d}}

\def\tensor#1{{\underline{\underline {#1}}}}

\def\bn{\begin{itemize}}
\def\en{\end{itemize}}

\begin{document}
\bibliographystyle{prsty}

\title{Universal Breakdown of Elasticity at the Onset of Material Failure}
\author{Craig Maloney$^{(1,2)}$}
\author{Ana\"el Lema\^{\i}tre$^{(1,3)}$}
\affiliation{
$^{(1)}$ Department of physics, University of California, Santa Barbara, California 93106, U.S.A.}
\affiliation{$^{(2)}$ Lawrence Livermore National Lab - CMS/MSTD, Livermore, California 94550, U.S.A.}
\affiliation{$^{(3)}$ L.M.D.H. - Universite Paris VI, UMR 7603,
4 place Jussieu, 75005  Paris - France}
\date{\today}

\begin{abstract}
We show that, in the athermal quasi-static deformation of amorphous materials,
the onset of failure is accompanied by universal scalings 
associated with a \emph{divergence} of elastic constants. 
A normal mode analysis of the non-affine elastic displacement field
allows us to clarify its relation to the zero-frequency mode at 
the onset of failure and to the crack-like pattern 
which results from the subsequent relaxation of energy.
\end{abstract}
%\pacs{62.20.Dc,62.20.Fe,62.25.+g,72.80.Ng}
\maketitle

\begin{figure}
\rotatebox{-90}{\resizebox{!}{.48\textwidth}{{\includegraphics{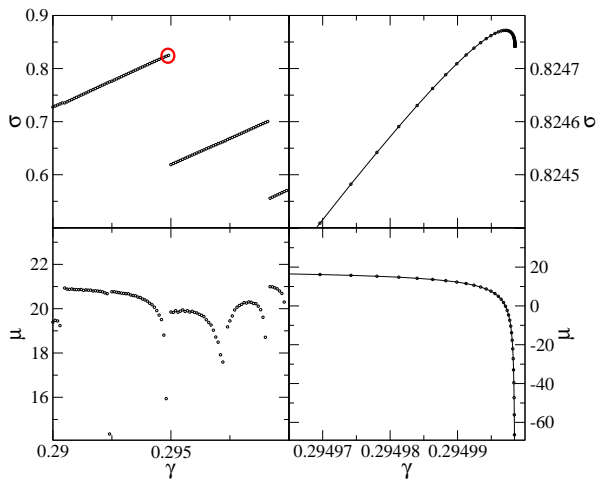}}}}
\caption{\label{fig:1}
Stress (top) and shear modulus (bottom) for a small strain interval about a strain of $.3$.
Left: fixed strain steps of size $10^{-4}$;
Right: convergence to the yield point (circled on the left)
with decreasing strain step.}
\end{figure}

Experiments on nanoindentation of metallic glasses~\cite{SchuhN03},
on granular materials~\cite{MillerOB96} and on foams~\cite{PrattD03},
demonstrate that at very low temperature and strain rates,
the microstructural mechanisms of deformation
involve highly intermittent stress fluctuations.
These fluctuations can be accessed in molecular dynamics simulations, 
but are best characterized numerically via
``exact'' implementation of a-thermal quasi-static deformation:
alternating elementary steps of affine deformation with energy 
relaxation~\cite{MaedaT78}
permits one to constrain the system to reside 
in a local energy minimum (inherent structure) at all times.
As illustrated in figure~\ref{fig:1}, macroscopic stress fluctuations 
arise from a series of reversible (elastic) branches corresponding to 
deformation-induced changes of local minima.
These branches are interrupted by sudden irreversible (plastic) events 
which occur when the inherent structure annihilates during
a collision with a saddle point.~\cite{MalandroL99}
These transitions constitute the most elementary mechanism of deformation 
and failure for disordered materials at low temperature.

Using this quasi-static protocol, recent studies of 
both elasticity~\cite{tanguy02} and plasticity~\cite{MalandroL99},
could identify important properties of elasto-plastic behavior which
arise solely from the geometrical structure of the potential energy landscape.
%have revealed striking properties which arise solely from the non-trivial 
%structure of the potential energy landscape.
Tanguy {\it et al\/}~\cite{tanguy02} have observed that, following
{\it reversible} (elastic) changes of the inherent structures,
molecules undergo large scale non-affine displacements.
They have shown these non-affine displacements to be
related to the breakdown of classical elasticity at small scales and 
to quantitative differences between measured Lam\'e constants 
and their Born approximation.
%Radja\"i and Roux have performed quasistatic simulations of 
%granular flow~\cite{RadjaiR01}.
Malandro and Lacks~\cite{MalandroL99} have shown that the destabilization of a minimum 
occurs through shear-induced collision with a saddle.
At the collision, a single normal mode sees its eigenvalue going to zero.
Building on this work, we studied the {\it irreversible}
(plastic) event following the disappearance of an inherent structure:
subsequent material deformation in search of a new minimum involves
non-local displacement fields --in the likeness of nascent cracks-- 
controlled by long-range elastic interactions.~\cite{MaloneyL04}

Several molecular displacement fields thus appear to be closely related to 
the geometrical structure of the potential energy landscape: 
(i) non-affine displacements along elastic
branches, (ii) the single normal mode controlling 
the annihilation of an inherent structure,
and (iii) the overall deformation occurring during an irreversible event.
In order to piece together a complete picture of elasto-plasticity at the nanoscale,
we need to understand the relation between these different fields and ask
how elastic behavior breaks down at the onset of failure.
It is thus a study of incipient plasticity 
--that is the onset of irreversible deformation-- that we wish to perform.
Here, the structural disorder is expected to control the onset of failure:
this situation is somehow opposite to homogeneous defect nucleation
in crystals,~\cite{li03} where failure
is controlled by Hill's continuum condition.~\cite{hill62}

We base our approach on exact microscopic expressions for the non-affine corrections 
to elasticity in disordered solids,~\cite{wallace72,lutsko88}
which have entirely been overlooked in recent works.
Here, we put such analytical tractations in perspective with the
recent numerical developments.  
We derive an exact formulation for the non-affine 
displacement fields, and construct a normal mode decomposition therefrom.
This analytical framework permits us to evidence that the 
lowest frequency normal mode dominates the non-affine elastic displacement field close 
to a plastic transition.  We then show that at any plastic transition point,
energy, stress, and the vibration frequencies display a singular, universal, 
behavior associated with a {\em divergence} of the elastic constants.
The normal mode analysis of the subsequent cascade shows that the low frequency 
modes active at incipient plasticity are superseded by long-range elastic 
interactions in the latter stages of an irreversible event.

%To this end, we have performed athermal, quasistatic simulations of a binary system of particles interacting through both a hertzian repulsive interaction and a Lennard-Jones type interaction.

%This framework allows us to obtain a series of analytical and numerical 
%results pertaining to two types issues: 
%the non-affine corrections to bulk elastic constants 
%in amorphous solids and the breakdown of elastic behavior at the onset 
%of material failure.
%Specifically, 
%(i) we clarify how non-affine contributions to the elastic constants 
%result from transfers of elastic energy to increasingly localized phonons,
%%in relation to a ``turbulent like'' cascade observed by Radjai and Roux.~\cite{radjai} 
%then (ii) we show that the lowest phonon mode dominates the 
%non-affine displacement field close to a plastic transition;
%this enables us to (iii) perform an analytical expansion for the non-affine field 
%and the elastic constants around a plastic event.
%From this expansion it results that at any plastic transition point,
%energy, stress, and the vibration frequencies display a singular, universal, 
%behavior associated to a {\em divergence} of the elastic constants.

%References:  Hutnik {\it etal}~\cite{hutnik93} use approach defined
%by Theodorou and Suter.~\cite{theodorou86a,theodorou86b}
%Lutsko~\cite{lutsko88} use affine transformation to define elastic tensor.

We consider in this work a molecular system in a periodic cell.
The geometry of the cell is determined by the matrix $h$ whose columns are
the Bravais vectors.~\cite{ray84,Ray85} 
The affine deformation of the cell between configurations $h_0$ and $h$ 
is characterized by the Green-St~Venant strain tensor, 
$\tensor\epsilon=
\frac{1}{2}\left((h_0^{-1})^T.h^T.h.h_0^{-1}-1\right)$,
which governs the elongation of a vector $\mathring{\vec x}\to\vec x$, as
$\vec x^2=\mathring{\vec x^2}+2\,\mathring{\vec x^T}.\tensor\epsilon.\mathring{\vec x}$.
As the energy functional generally depends only on the set of 
interparticle distances, 
it can be parameterized as ${\cal U}(\{\mathring{\vec r_i}\},\tensor\epsilon)$ 
where $\{\mathring{\vec r_i}\}$ are the positions of the particles
in a {\it reference} cell.~\cite{barron65,wallace72} 
Varying, $\tensor\epsilon$ for fixed $\{\mathring{\vec r_i}\}$ 
corresponds to an affine displacement of the molecules in real space.

To start, 
%we introduce analytical expressions for the elastic constants and non-affine displacement fields. 
%To derive this expression,
let's contemplate more closely the athermal, quasi-static algorithm.
Deformation, $\tensor\epsilon(\gamma)$,
is enforced by moving the Bravais axes of the periodic cell;
$\gamma$ is introduced as rescaled coordinate which measures the amount of
deformation from some reference state.
In practice $\tensor\epsilon(\gamma)$ corresponds to either pure shear
or pure compression.
Formally, the limit $h_{0}=h$ (or $\gamma\to0$) is often appropriate to define stresses
and elastic constant around a (possibly stressed) reference state.
%The algorithm is designed to follow deformation-induced changes 
%of a local minimum in the potential energy landscape.
Once a choice of $h_0$ is made, the algorithm tracks in the reference cell
a trajectory $\{\mathring{\vec r_i}\}(\gamma)$, which is implicitly
defined by demanding that the system remain in 
mechanical equilibrium:~\cite{wallace72,lutsko88}
\begin{equation}
\label{eqn:mechanical}
% \forall\,i\quad,\quad\quad\vec F_i\equiv
%\frac{\partial{\cal U}}{\partial\vec r_i}\Big|_{\tensor\epsilon} 
%(\{\vec r_j\},\tensor\epsilon) = \vec 0 
%\quad.
 \forall\,i\quad,\quad\quad\vec F_i\equiv
\frac{\partial{\cal U}}{\partial\mathring{\vec r_i}}\Big|_{\gamma} 
(\{\mathring{\vec r_j}\},\gamma) = \vec 0 
\quad.
\end{equation}
Starting at mechanical equilibrium at $\gamma=0$, 
$\{\mathring{\vec r_i}(\gamma)\}$ is a continuous function of $\gamma$ 
on some interval $[0,\gamma_c]$. 
At $\gamma_c$, the local minimum collides with a saddle point.~\cite{MalandroL99}

An equation of motion for $\mathring{\bf\vec r}=\{\mathring{\vec r_i}(\gamma)\}$ 
is obtained by
derivation of~(\ref{eqn:mechanical}) with respect to $\gamma$.
%\begin{equation}
%\forall\,i\quad,\quad\quad\sum_j\frac{\partial^2{\cal U}}{\partial\vec r_i\partial\vec r_j}\,.\,\frac{\d\vec r_j}{\d\gamma}
%+\frac{\partial^2{\cal U}}{\partial\vec r_i\partial\gamma} = \vec 0
%\end{equation}
Denoting,
${\bf\cal H} = \left(\frac{\partial^2{\cal U}}{\partial\mathring{\vec r_i}\partial\mathring{\vec r_j}}\right)$,
${\bf \vec \Xi} = \left(\frac{\partial^2{\cal U}}{\partial\mathring{\vec r_i}\partial\gamma}\right)$,
and
${\bf \vec \Xi}_{\alpha\beta} = \left(\frac{\partial^2{\cal U}}{\partial\mathring{\vec r_i}\partial\epsilon_{\alpha\beta}}\right)$, we find:
%the phase-space trajectory ${\bf\vec r}=\{r_i(\gamma)\}$ verifies:
\begin{equation}
\label{eqn:naf}
\frac{\d{\mathring{\bf\vec r}}}{\d\gamma}=
-{\bf\cal H}^{-1}.{\bf \vec \Xi}
=
-{\bf\cal H}^{-1}.\sum_{\alpha\beta}{\bf \vec \Xi}_{\alpha\beta}
\,\frac{\d\epsilon_{\alpha\beta}}{\d \gamma}
\quad.
\end{equation}
This relation holds for any $\gamma\in[0,\gamma_c]$.
In the limit $h\to h_0$, ${\cal H}$ is the Dynamical Matrix.
To invert ${\cal H}$, translation modes must be eliminated 
by fixing the position of a molecule.
$\d\mathring{\bf\vec r}/\d\gamma$ is a rescaled ``velocity'' of molecules 
in quasi-static deformation.
It defines the direction (in tangent space) of the non-affine displacement 
field observed by Tanguy {\it et al\/} and can be directly evaluated 
by solving equation~(\ref{eqn:naf})
without resorting to quadruple precision minimization.~\cite{tanguy02}

\begin{figure}
\resizebox{!}{.22\textwidth}{{\includegraphics{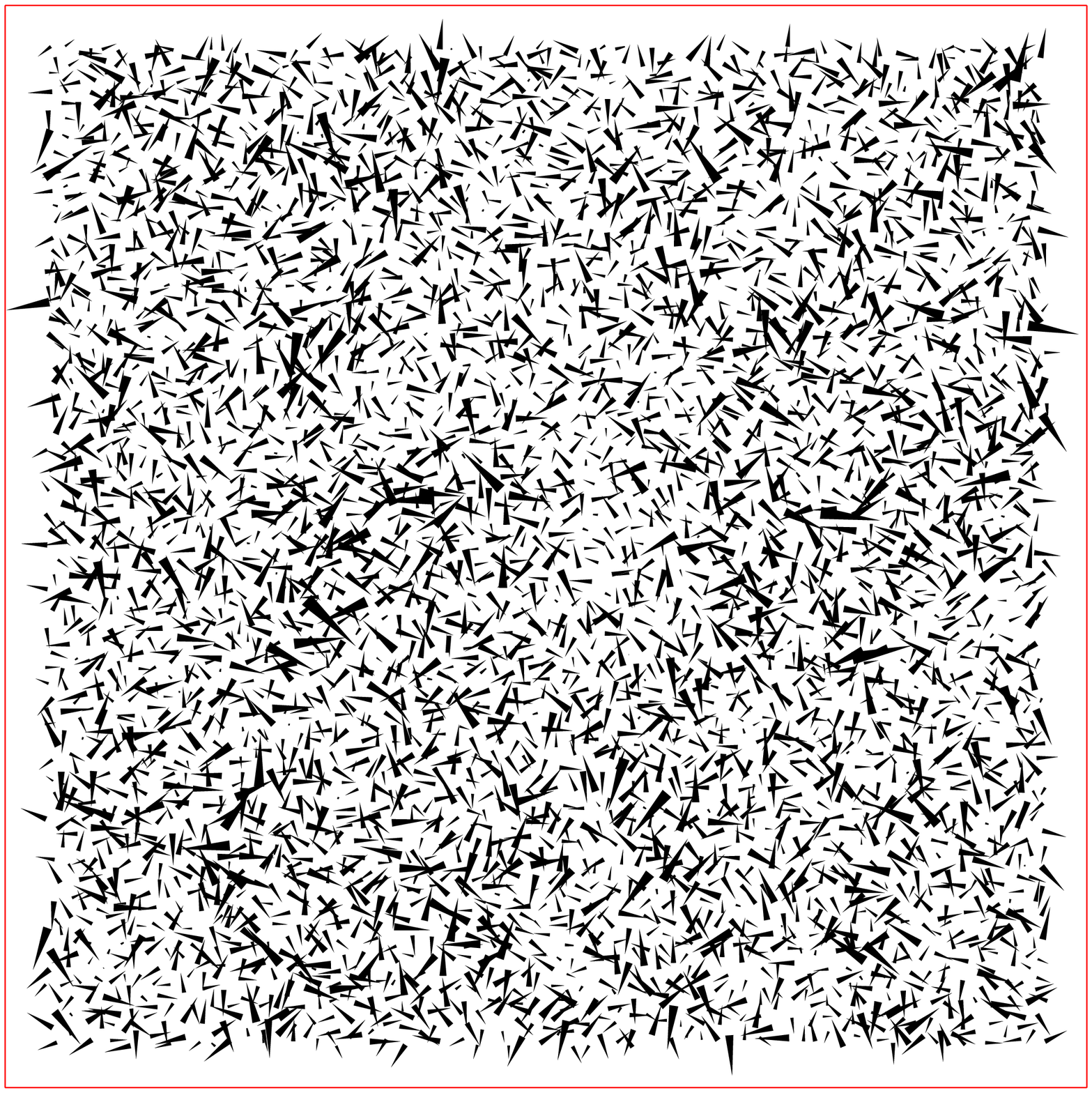}}}
\resizebox{!}{.22\textwidth}{{\includegraphics{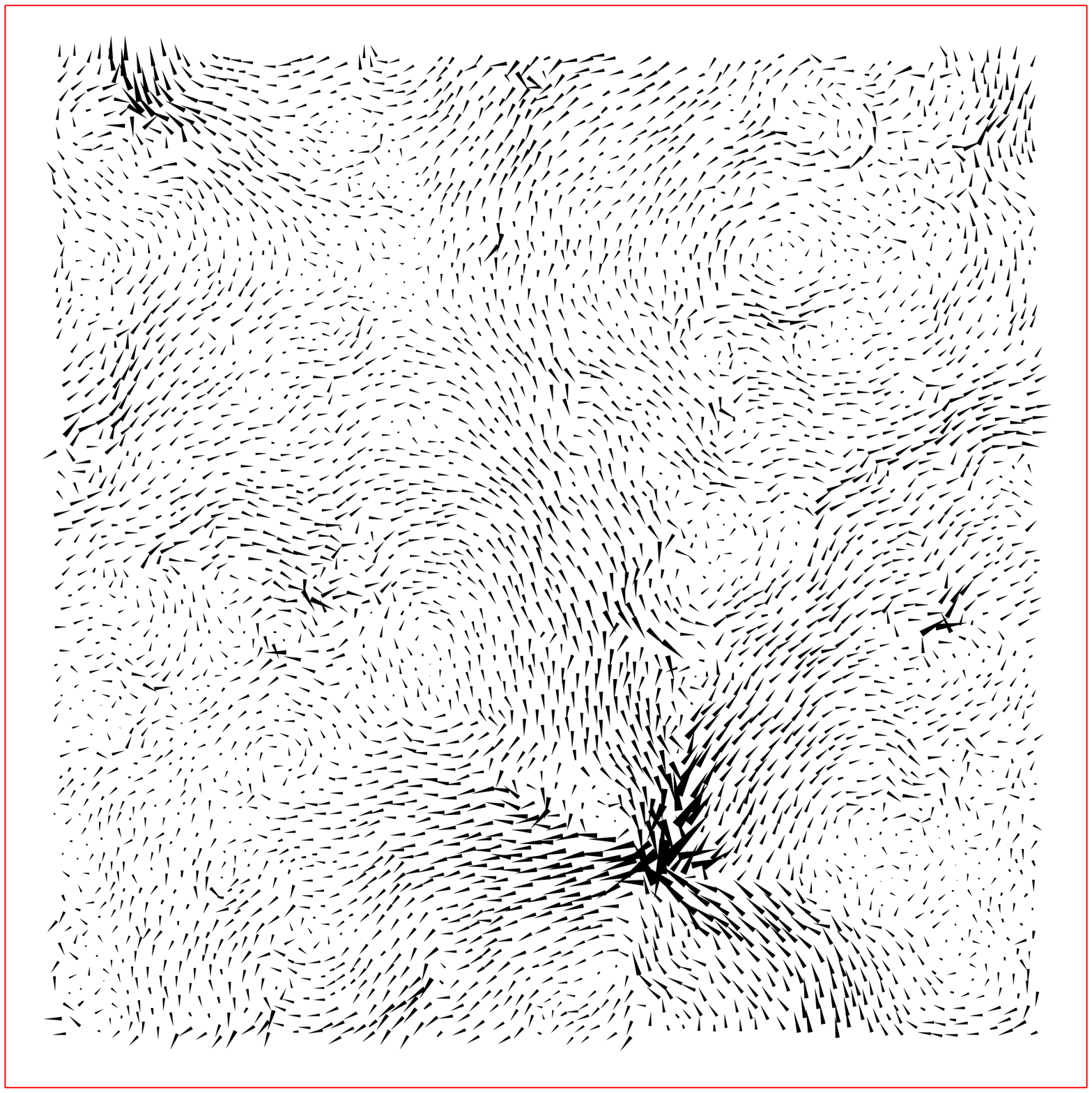}}}
\caption{\label{fig:fields}
Left: The force response to simple shear, 
$\vec{\Xi}$, 
at a strain configuration, 
$\gamma=0.2945$, 
or $\gamma_c-\gamma\sim 10^{-4}$.  
Right: The non-affine velocity (or ``displacement'') field, 
$\frac{\d{\bf\vec r}}{\d\gamma}$ 
for the same state as shown on the left.}
\end{figure}

%We illustrate these displacement field on a numerical model of....
%in a pure shear geometry.
Here, we illustrate these ideas with numerical simulations of a two-dimensional
bidisperse mixture of particles interacting through a shifted
Lennard-Jones potential.~\cite{tanguy02}
Particle sizes $r_{S}=r_{L}{\sin{\frac{\pi}{10}}}/{\sin{\frac{\pi}{5}}}$
and a number ratio $N_{L}/N_{S}=\frac{1+\sqrt{5}}{4}$ 
are used to prevent crystallization;  
the simulation cell is 50$\times r_{L}$ in length.
We have also performed simulations on Hertzian spheres
to check that it yielded results consistent with those presented here.
Typical patterns of the fields ${\bf \vec \Xi}$ and $\d\mathring{\bf\vec r}/\d\gamma$ 
in (steady) simple shear deformation are shown figure~\ref{fig:fields}: the apparent
small scale randomness of the vector ${\bf \vec \Xi}$ is in sharp contrast
with the large vortex-like structures displayed by the non-affine ``velocity''
field $\d\mathring{\bf\vec r}/\d\gamma$.
To understand the randomness of ${\bf \vec \Xi}$, note that
${\vec \Xi}_i=\partial\vec F_i/\partial \gamma$ is the force response
on molecule $i$ after an elementary affine deformation of the system:
it only depends on the configuration of the molecules with which molecule $i$ interacts,
hence is an $\tensor\epsilon$-dependent measure of the local disorder 
of molecular configurations.
We checked that spatial correlations decay very fast in the
field ${\bf \vec \Xi}$:
in the following discussion, 
the short-range randomness of the field ${\bf \vec \Xi}$
allows us to interpret it as noise.

An analytical expression for the bulk elastic constants 
derives along similar lines.~\cite{wallace72,lutsko88}
The first derivative of the potential with respect to the components
of $\tensor\epsilon$ defines the thermodynamic stress, $\tensor t$:
$
t_{\alpha\beta} 
= \frac{1}{V_0}\,\frac{\d\,{\cal U}}{\d\,\epsilon_{\alpha\beta}}
= \frac{1}{V_0}\,\frac{\partial\,{\cal U}}{\partial\,\epsilon_{\alpha\beta}}
$
The total derivative indicates derivation while preserving mechanical equilibrium,
the second equality results from equation~(\ref{eqn:mechanical}),
and $V_0$ is the volume of the simulation cell.
The second (total) derivative of the energy gives the elastic constants,~\cite{barron65}
\begin{equation}
C_{\alpha\beta\chi\sigma}
%=\frac{\d^2 {\cal U}}{\d\epsilon_{\alpha\beta}\d\epsilon_{\chi\sigma}} 
= \frac{1}{V_0}\,
\left(
\frac{\partial^2{\cal U}}{\partial\epsilon_{\alpha\beta}\partial\epsilon_{\chi\sigma}}
+\sum_{j}\frac{\partial^2{\cal U}}
{\partial\mathring{\vec r_i}\partial\epsilon_{\alpha\beta}}\,.\,\frac{\d\mathring{\vec r_i}}{\d\epsilon_{\chi\sigma}}
\right)
\quad.
\end{equation}
We recognize the first term as being the Born approximation $C^{\rm Born}_{\alpha\beta\chi\sigma}$.
The second term accounts for the non-affine corrections, and reads:
$\widetilde C_{\alpha\beta\chi\sigma}
=-\frac{1}{V_0} {\bf\vec\Xi}_{\alpha\beta}.{\bf\cal H}^{-1}.{\bf\vec\Xi}_{\chi\sigma} 
$.
Similarly, the second derivatives of the energy, 
following any generic deformation $\tensor\epsilon(\gamma)$, can be written as:
\begin{equation}
\frac{\d^2 U}{\d\gamma^2}
%= {\bf\vec\Xi}_{\alpha\beta}.\frac{\d{\bf\vec r}}{\d\epsilon_{\chi\sigma}} 
= \frac{\partial^2 U}{\partial\gamma^2}
- {\bf\vec\Xi}.{\bf\cal H}^{-1}.{\bf\vec\Xi} 
%=
% - \frac{\d{\bf\vec r}}{\d\epsilon_{\chi\sigma}}.{\bf\cal H}.
%\frac{\d{\bf\vec r}}{\d\epsilon_{\chi\sigma}} 
\quad.
\label{eqn:xidef}
\end{equation}

For an isotropic material, the elastic constants can be written:
$C_{\alpha\beta\chi\sigma}
=
\lambda\,\delta_{\alpha\beta}\,\delta_{\chi\sigma}
+\mu\,(\delta_{\alpha\chi}\,\delta_{\beta\sigma}+\delta_{\alpha\sigma}\,\delta_{\beta\chi})
$, which define the Lam\'e constants, $\lambda$ and $\mu$. 
In order to estimate these constants, it is not necessary to evaluate all the 
components of the tensor ${\tensor{\bf\vec\Xi}}=({\bf\vec\Xi}_{\alpha\beta})$, 
but only two of its projections
${\bf\vec\Xi}$, {\it e.g.} for pure shear and pure compression
%--this corresponds to monitoring the response of the system to two types of tests--
and use equation~(\ref{eqn:xidef}).
In equation~(\ref{eqn:xidef}) the correction to the Born 
term is negative definite: quantities such as the shear modulus, $\mu$, 
or the compressibility, $K=\lambda + \mu$, are necessarily 
smaller than the Born term, while this is not necessarily true
of $\lambda=K-\mu$ alone as it does not, by itself, correspond to any 
realizable mode of deformation.
This is consistent with the numerical observations
%FIXME FIND THIS CITATION!!!
%experimentally by Lossert etal~\cite{lossert} in colloidal systems and
by Tanguy {\it et al\/}~\cite{tanguy02} in Lennard-Jones systems.

%Knowing ${\bf\vec\Xi}$ and the non-affine displacement 
%direction $\d{\bf\vec r}/\d\gamma$, 
%we can calculate the elastic constants in a single step.

Next, we perform a normal mode analysis of the fields ${\bf\vec\Xi}$.
Denoting ${\bf\vec\Psi}_p$ the eigenvectors
of the Dynamical Matrix (normal modes),
and $\lambda_p$ the associated eigenvalues,
the vector ${\bf \vec\Xi}$ can be decomposed as:
%\begin{equation}
${\bf \vec\Xi} = \sum_p \xi_p\,{\bf\vec\Psi}_p,$
%\end{equation}
with $\xi_p={\bf \vec\Xi}.{\bf\vec\Psi}_p$.
(If ${\bf \vec\Xi}$ is a random field, the variables $\xi_p$
are random.) 
From this decomposition, expressions can be obtained for
the non-affine direction and for the non-affine contribution to elasticity:
\begin{equation}
\label{eqn:munormal}
\frac{\d\mathring{\bf\vec r}}{\d\gamma}=
-\sum_p \frac{\xi_p}{\lambda_p}\,{\bf\vec\Psi}_p\quad\mbox{and}\quad 
C_{\tensor\epsilon} = -\sum_p \frac{\xi_p^2}{\lambda_p}
\quad.
\end{equation}

\begin{figure}
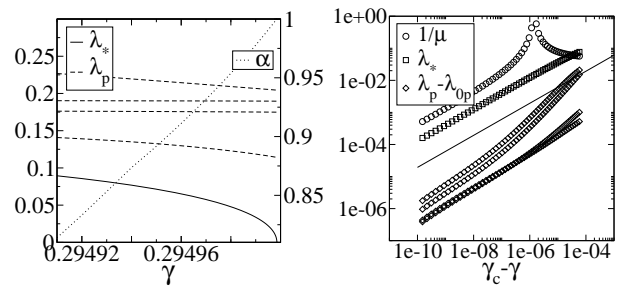

%\rotatebox{-90}{\resizebox{!}{.23\textwidth}{{\includegraphics{fig3a.ps}}}}
%\resizebox{!}{.20\textwidth}{{\includegraphics{fig3a.ps}}}
%\rotatebox{-90}{\resizebox{!}{.20\textwidth}{{\includegraphics{fig3b.ps}}}}
\includegraphics[width=.23\textwidth]{fig3a.eps}
\includegraphics[width=.21\textwidth]{fig3b.eps}
\caption{\label{fig:convergence}
Left: relative participation of the lowest normal mode in the non-affine elastic displacement field, 
$\alpha^*\doteq\left(\xi^{*}/\lambda^{*}\right)^{2}/\sum_{p}\left(\xi_{p}/\lambda_{p}\right)^2$ 
(dotted); lowest eigenvalue of the dynamical matrix (solid); next several eigenvalues (dashed).  b) In log-log scale
(as a guide to the eye, the thick black line is $\sqrt{\gamma_c-\gamma}$): 
$1/\mu$ (circles); lowest eigenvalue (squares); next several eigenvalues minus their terminal values (diamonds).}
\end{figure}

We now concentrate on the behavior of the shear modulus at incipient plasticity, 
as shown in figure~\ref{fig:1} and~\ref{fig:convergence}.
Malandro and Lacks have shown numerically that at the onset of a plastic event 
a single eigenfrequency goes to zero.~\cite{MalandroL99}
We denote ${\bf\vec\Psi}^*(\gamma)$ 
the first non-zero normal mode; in two dimensions, 
it is the third in the spectrum.
Close to failure ($\gamma\to\gamma_c$), $\lambda^*(\gamma)\to 0$, 
hence ${\bf\vec\Psi}^*(\gamma_c)$ must dominate the non-affine 
direction $\d\mathring{\bf\vec r}/\d\gamma$.
This is true if and only if the quantity 
$\xi^*(\gamma)={\bf\vec\Psi}^*(\gamma).{\bf\vec\Xi}(\gamma)$ 
does not vanish at the yield point.

In order to check this scenario,
we have performed numerical simulations of the same 2D 
binary mixture described above. 
We note that, on approaching a plastic event, caution must be taken 
to correctly minimize the energy without using a quadratic approximation.
We observe that (i) the mode ${\bf\vec\Psi}^*$ is localized
close to failure 
%(consistently with~\cite{MalandroL99})
while (ii) ${\bf\vec\Xi}(\gamma)\to{\bf\vec\Xi}(\gamma_c)$ 
remains noisy and weakly correlated with the normal modes.
As a consequence of this observation, $\xi^*(\gamma)$ has 
a random, but typically non-zero limit when $\gamma\to\gamma_c$.
The non-affine field is dominated by 
$-\xi^*(\gamma_c)/\lambda^*(\gamma)\times{\bf\vec\Psi}^*(\gamma_c)$ 
and the non-affine correction to elasticity by
$\widetilde \mu\sim-(\xi^*)^2/\lambda^*$ which diverges toward $-\infty$
(see figure~\ref{fig:1}).
In contrast, the Born term--which depends on the pair-correlation only--
does not present any divergence.
Since $\mu=\mu^{\rm Born}+\widetilde \mu$, 
on approaching failure, the system reaches a point $\gamma_0<\gamma_c$
at which $\widetilde \mu=-\mu^{\rm Born}$, whence $\mu$ vanishes.
For $\gamma\in[\gamma_0,\gamma_c]$, the shear stress is a decreasing 
function of $\gamma$:
the material is unstable to any constant applied stress.
This region is accessible to us because
deformation--and not stress--is prescribed.

In order to understand more specifically how the elastic 
constants behave close to $\gamma_c$, 
let us consider the functions $\lambda_p(\gamma)$, which are continuous
on a small interval close to $\gamma_c$ (the second derivatives
of the potential are supposed to be regular).
Close to the yield point, $\gamma_c$, the deformation is dominated by the lowest normal
mode: $\mathring{\bf\vec r}(\gamma)-\mathring{\bf\vec r}(\gamma_c)
\sim x(\gamma)\,{\bf\vec\Psi}^*(\gamma_c)$
%, with $x(\gamma)= (\mathring{\bf\vec r}(\gamma)-\mathring{\bf\vec r}(\gamma_c)).{\bf\vec\Psi}^*(\gamma_c)$
.
(We project the deformation on the mode ${\bf\vec\Psi}^*$ {\it at}
the yield point.)
From this relation and~(\ref{eqn:munormal}), we obtain the dominant contribution:
$\d x/\d \gamma \sim -\xi^*(\gamma_c)/\lambda^*(\gamma)$.
The coordinate $x$ measures a true displacement in configuration space:
we expect that no singular behavior occurs in this rescaled coordinate whence,
$\lambda^*(x)$ should vanish regularly, $\lambda^*(x)\sim a x$ close to $x=0$.
From this assumption, it results:
$x(\gamma)\sim\sqrt{2\xi^*(\gamma_c)(\gamma_c-\gamma)/a}$. This relation
controls entirely the behavior of all observables when approaching the yield point:
any observable $A$ which behaves regularly as a function of $x$ (any regular
function of molecular configurations) ``accelerates'' close to the yield point:
$\d A/\d \gamma \sim 1/\sqrt{\gamma_c-\gamma}$.
In particular, we obtain, 
$\d \mathring{\bf\vec r}/\d \gamma\sim{\bf\vec\Psi}_p/\sqrt{\gamma_c-\gamma}$,
and $\lambda^*(\gamma) = \sqrt{2a\xi^*(\gamma_c-\gamma)}$,
whence, $\widetilde\mu\sim-(\xi^*)^{3/2}/\sqrt{2a(\gamma_c-\gamma)}$.
We could observe these scalings numerically by a careful approach to the yield point,
as shown figure~\ref{fig:convergence}.
A similar divergence is observed for the compression modulus, but with a different
prefactor, determined by the normal mode decomposition 
 of the ${\bf\vec\Xi}$ field associated with pure compression.

\begin{figure}
\resizebox{!}{.22\textwidth}{{\includegraphics{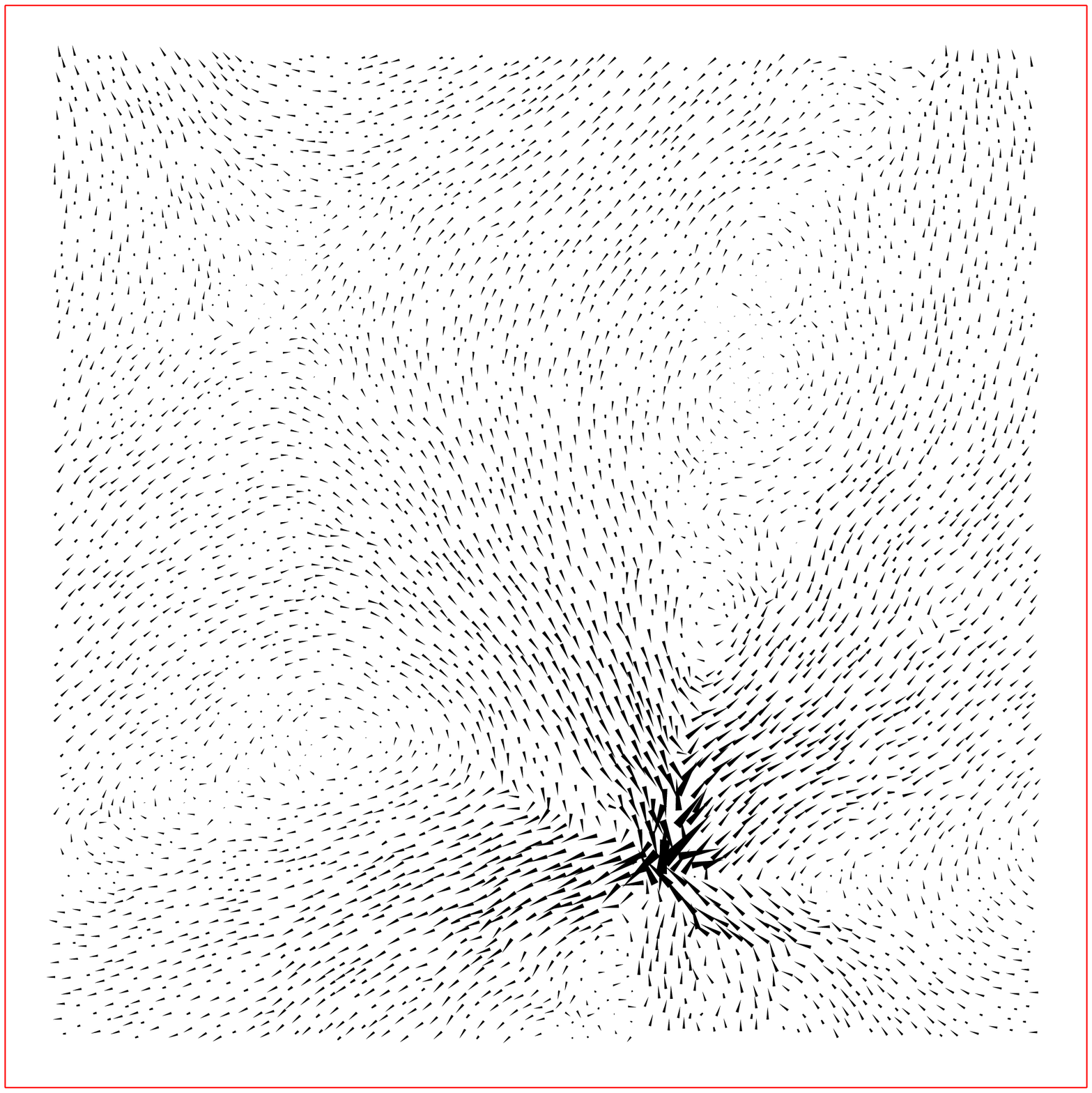}}}
\resizebox{!}{.22\textwidth}{{\includegraphics{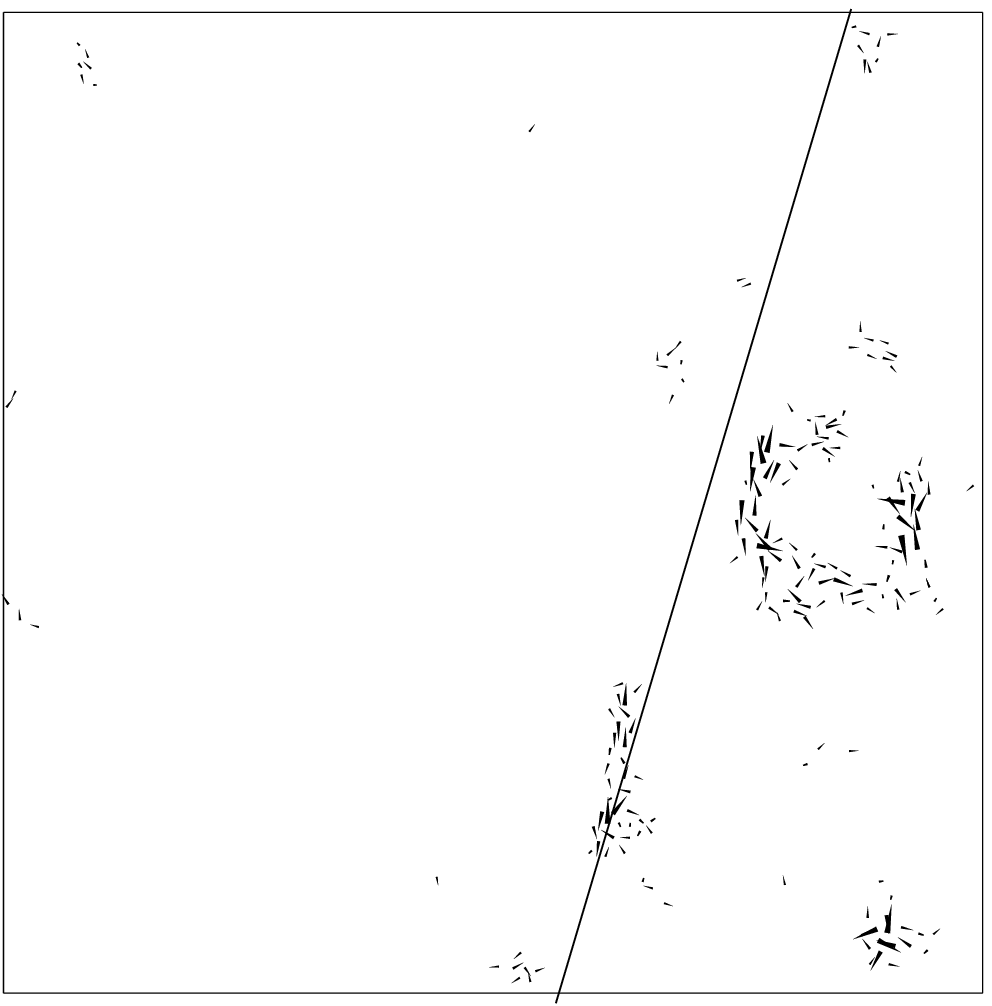}}}
\caption{\label{fig:fields2}
Left: Non-affine elastic displacement field at a distance $\gamma_c-\gamma\sim 10^{-10}$ 
from the transition.  Note the quadrupolar alignment with the direction of applied strain.  
%The response is completely dominated by the single lowest eigenmode at this strain.
%Note the dominant quadrupolar displacement field which has its tensile and compressive axes aligned with the tensile and compressive directions of the imposed simple shear.
Right: The local relative displacement field (the displacement of each particle measured with respect to the average displacement of its neighbors) which is incurred after the entire plastic cascade.  The solid line is a guide to the eye oriented along the oblique Bravais axis.
}
\end{figure}

\begin{figure}
\includegraphics[width=.4\textwidth]{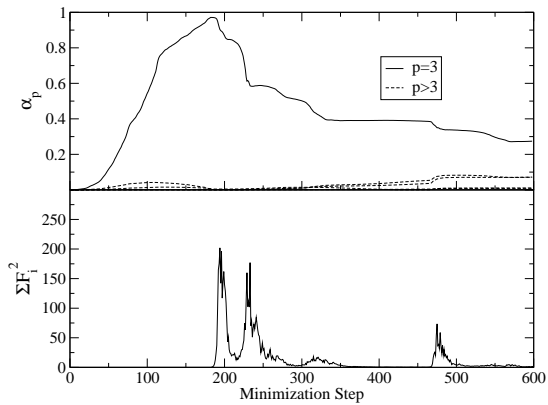}
\caption{\label{fig:cascade}
Evolution of the displacement field during the irreversible cascade
corresponding to the event circled in figure~\ref{fig:1}.
Top: contribution $\alpha^*$ (solid line) of the critical mode 
and $\alpha_p(t)$ (dashed line) of the next five modes to the
displacement field. 
Bottom: sum of the squares of the forces on the particles.}
\end{figure}

We now turn to the overall plastic event following failure 
(see figure~\ref{fig:fields2} 
and~\ref{fig:cascade}).
We have already shown, in similar atomistic systems, that any single
plastic event involves a cascade of local rearrangements.~\cite{MaloneyL04}
Our preceding work suggested that the overall plastic event
was controlled by 
%macroscopic elasticity, 
long range elastic interactions
and differed from the displacement
fields which dominate the onset of failure.
Our present normal mode decomposition allows us to gain more insight into this process.
%To this end, we study the projection on ${\bf\vec\Psi}^*$ of the dynamics 
%of $\mathring{\bf\vec r}(t)$ during the relaxation corresponding to 
%a full irreversible cascade ($t$ indicates the rescaled time along steepest descent).
Writing $\mathring{\bf\vec r}(t)
-\mathring{\bf\vec r}(0)=\sum_p \Delta\xi_p(t)\,{\bf\vec\Psi}_{p}(\gamma_c)$,
we extract the quantities 
$\alpha_p\doteq(\Delta\xi_p(t))^2/\sum_p(\Delta\xi_p(t))^2$, which are shown 
figure~\ref{fig:cascade} for the lowest frequency modes.
To trigger the relaxation, we shear the system forward by a small
amount of shear, $\gamma-\gamma_c\sim 10^{-5}$.
The initial affine displacement serves as a perturbation and
projects randomly on the normal modes, 
whence the contributions $\alpha_p$ start around zero. 
%Then the zero frequency mode grows, as the system attempts to join the 
%Then the zero frequency mode grows, as the system relaxes toward the 
%recently annihilated minimum. 
We observe that
(i) the initiation of the cascade is clearly dominated by the critical mode
(ii) this effect suddenly stops before reaching the first peak in $\sum_i F_i^2$
(this peak correspond to the first inflection point of energy vs. minimization step)
(iii) the subsequent displacement appears to be random, when projected on the lowest
part of the spectrum, indicating that low frequency normal modes are irrelevant 
for the latter stages of plastic failure.
The system escapes its initial inherent structure 
in the direction of the lowest normal mode, whereas, 
in the late stages of a plastic event 
the emergence of long-range elastic interactions supersedes
the low energy modes responsible for the onset of failure.

To conclude, we wish to stress that we expect our analysis to apply to 
many materials, to modes of deformation other than uniform shear, 
and to several experimental protocols at the nanoscale, 
including nanoindentation studies.
% or material deformation 
%induced by an AFM tip. 
Knowing the detailed behavior of elastic
constants at incipient plasticity opens the route toward a possible control of
material deformation at the nanometer scale.
%as much as it unveils 
%the fundamental origin of plasticity.

%We believe that present work allows to draw a very detailed picture 
%of the transition between elastic branches and plastic events, and 
%gain insight on the essential mechanism control plastic deformation.
%Such a picture has never been proposed 
%its effect greatly diminishes over time; by the end of the 
%cascade, it accounts for only 20 percent of the entire displacement
%field.  Other more complicated patterns emerge.  For instance, the
%$p=6$ phonon does not participate much in the cascade until a late 
%stage, after which it accounts for roughly the same portion of the
%deformation as does the critical phonon, while the $p=4$ and $p=5$
%essentially do nothing throughout the cascade.  These complex
%interactions during the cascade cannot be 
%fully understood through an analysis of the
%potential energy surface at the critical point.

This work was partially supported under the auspices of the U.S. Department of Energy by the University of California, Lawrence Livermore National Laboratory under Contract No. W-7405-Eng-48, by the NSF under grants DMR00-80034 and DMR-9813752, by the W. M. Keck Foundation, and EPRI/DoD through the Program on Interactive Complex Networks. CM would like
to acknowledge the guidance and support of V.~V. Bulatov and J.~S. Langer and the hospitality of LLNL University Relations.

%\bibliography{../thermo}

\begin{thebibliography}{10}

\bibitem{SchuhN03}
C.~A. Schuh and T.~G. Nieh, Acta Mater. {\bf 51},  87  (2003).

\bibitem{MillerOB96}
B. Miller, C. OHern, and R.~P. Behringer, Phys. Rev. Lett. {\bf 77},  3110
  (1996).

\bibitem{PrattD03}
E. Pratt and M. Dennin, Phys. Rev. E {\bf 67},   051402   (2003).

\bibitem{MaedaT78}
K. Maeda and S. Takeuchi, J Phys-F-Metal Phys {\bf 8},  L283  (1978).

\bibitem{MalandroL99}
D.~L. Malandro and D.~J. Lacks, J. Chem. Phys. {\bf 110},  4593  (1999).

\bibitem{tanguy02}
A. Tanguy, J.~P. Wittmer, F. Leonforte, and J.-L. Barrat, Phys. Rev. B {\bf
  66},  174205  (2002).

\bibitem{MaloneyL04}
C. Maloney and A. Lema\^{\i}tre, cond-mat/0402148  .

\bibitem{li03}
J. Li {\it et~al.}, Nature {\bf 418},  307  (2003).

\bibitem{hill62}
R. Hill, J. Mech. Phys. Sol. {\bf 10},  1  (1962).

\bibitem{wallace72}
D. Wallace, {\em Thermodynamics of Crystals} (Wiley, New York, 1972).

\bibitem{lutsko88}
J.~F. Lutsko, J. Appl. Phys. {\bf 64},  1152  (1988).

\bibitem{ray84}
J. Ray and A. Rahman, J. Chem. Phys. {\bf 80},  4423  (1984).

\bibitem{Ray85}
J.~R. Ray, M.~C. Moody and A. Rahman, Phys. Rev. B {\bf 32},  733  (1985).

\bibitem{barron65}
T. Barron and M. Klein, proc. Phys. Soc. {\bf 85},  523  (1965).

\end{thebibliography}

\end{document}